\documentclass[acmsmall,screen]{acmart}
\AtBeginDocument{%
  \providecommand\BibTeX{{%
    \normalfont B\kern-0.5em{\scshape i\kern-0.25em b}\kern-0.8em\TeX}}}
    
\setcopyright{acmcopyright}
\acmJournal{TOSEM}
\acmYear{2024} \acmVolume{1} \acmNumber{1} \acmArticle{1} \acmMonth{7} \acmPrice{15.00}

\acmPrice{15.00}
\acmISBN{978-1-4503-XXXX-X/18/06}

\usepackage{xspace}
\usepackage{graphicx}
\usepackage{url}
\usepackage{multirow}
\usepackage{makecell}
\usepackage{colortbl}
\usepackage{tcolorbox}

\newcommand{\colorcell}{\cellcolor[rgb]{ .851,  .851,  .851}}

\usepackage[framemethod=tikz]{mdframed}
\newcounter{finding}
\numberwithin{finding}{section}

\newcommand{\finding}[1]{\refstepcounter{finding}
  \vspace{2.3mm}
 \begin{mdframed}[linecolor=gray,roundcorner=12pt,backgroundcolor=gray!15,linewidth=3pt,innerleftmargin=2pt, leftmargin=0cm,rightmargin=0cm,topline=false,bottomline=false,rightline = false]
 
  \textbf{Ans. to } #1
 \end{mdframed}
 \vspace{2.3mm}
}

\usepackage{enumitem}

\begin{document}

\title{Bias Behind the Wheel: Fairness Testing of Autonomous Driving Systems}


\author{Xinyue Li}
\email{xinyueli@stu.pku.edu.cn}
\affiliation{%
  \institution{Peking University}
  \country{China}
}

\author{Zhenpeng Chen}\authornote{Corresponding author: Zhenpeng Chen (zhenpeng.chen@ntu.edu.sg).}
\email{zhenpeng.chen@ntu.edu.sg}
\affiliation{%
  \institution{Nanyang Technological University}
  \country{Singapore}
}

\author{Jie M. Zhang}
\email{jie.zhang@kcl.ac.uk}
\affiliation{%
  \institution{King's College London}
  \country{United Kingdom}
}

\author{Federica Sarro}
\email{f.sarro@ucl.ac.uk}
\affiliation{%
  \institution{University College London}
  \country{United Kingdom}
}

\author{Ying Zhang}
\email{ying.zhang@pku.edu.cn}
\affiliation{%
  \institution{Peking University}
  \country{China}
}

\author{Xuanzhe Liu}
\email{liuxuanzhe@pku.edu.cn}
\affiliation{%
  \institution{Peking University}
  \country{China}
}
\renewcommand{\shortauthors}{Li et al.}
\begin{abstract}
This paper conducts fairness testing of automated pedestrian detection, a crucial but under-explored issue in autonomous driving systems. We evaluate eight state-of-the-art deep learning-based pedestrian detectors across demographic groups on large-scale real-world datasets. To enable thorough fairness testing, we provide extensive annotations for the datasets, resulting in 8,311 images with 16,070 gender labels, 20,115 age labels, and 3,513 skin tone labels.
Our findings reveal significant fairness issues, particularly related to age. The proportion of undetected children is 20.14\% higher compared to adults. Furthermore, we explore how various driving scenarios affect the fairness of pedestrian detectors. We find that pedestrian detectors demonstrate significant gender biases during night time, potentially exacerbating the prevalent societal issue of female safety
concerns during nighttime out. Moreover, we observe that pedestrian detectors can demonstrate both enhanced fairness and superior performance under specific driving conditions, which challenges the fairness-performance trade-off theory widely acknowledged in the fairness literature. We publicly release the code, data, and results to support future research on fairness in autonomous driving.
\end{abstract}



\maketitle

\section{Introduction}\label{intro}
Autonomous driving systems are on track to become the predominant mode of transportation in the future \cite{icse0001WCT20}. However, these systems are susceptible to software bugs \cite{GarciaF0AXC20}, which can potentially result in severe injuries or even fatalities for both pedestrians and passengers. The unfortunate incident in 2018 involving an autonomous vehicle from Uber serves as a stark reminder of these risks \cite{caselink1}. Given the safety-critical nature of autonomous driving systems, they have garnered substantial attention from the software testing community \cite{tseZhangHML22}.

Extensive research efforts have been devoted to the testing of autonomous driving systems. For example, Tian et al. \cite{10.1145/3180155.3180220} introduced DeepTest, which applies image transformation to simulate potential camera noise in autonomous driving scenarios. Zhang et al. \cite{10.1145/3238147.3238187} developed DeepRoad, a Generative Adversarial Network (GAN)-based approach that generates test images from real-world driving scenes. Zhou et al. \cite{10.1145/3377811.3380422} proposed DeepBillboard, a system for generating adversarial billboards to induce potential steering errors in autonomous vehicles.

Although significant testing efforts have been made, to the best of our knowledge, the study of fairness testing for autonomous driving systems remains under investigation in the literature.
From the Software Engineering (SE) perspective, fairness is considered a non-functional software property, making it an important subject for testing \cite{tseZhangHML22,tosemSunCZH24}. Fairness testing, as an emerging domain within software testing, seeks to uncover fairness issues in software systems~\cite{chen2022fairness}. 

Fairness issues in autonomous driving systems, such as a higher accuracy in detecting male pedestrians compared to females, can perpetuate discriminatory outcomes and unequal treatment based on gender. This can result in harm to individuals belonging to marginalized groups, further exacerbating existing social inequalities. Therefore, it is crucial to prioritize fairness testing in autonomous driving systems.

To fill the gap, we conduct fairness testing of eight state-of-the-art Deep Learning (DL)-based pedestrian detectors that have been extensively studied in the research community. Our main focus is to quantitatively assess performance disparities in these detectors across diverse demographic groups, which are widely recognized as group fairness issues \cite{sigsoftChenZSH22}. To enable fairness testing, we manually enrich four widely-adopted real-world datasets with gender, age, and skin tone labels, resulting in a collection of 8,311 real-world images annotated with 16,070 gender labels, 20,115 age labels, and 3,513 skin tone labels. Using these labeled datasets, we assess the group fairness of existing pedestrian detectors and also explore how commonly-studied driving scenarios (including various brightness, contrast, and weather conditions) impact the fairness of these detectors.

Our study reveals the following findings:
\textbf{(1)} Overall, state-of-the-art pedestrian detectors exhibit significant bias regarding age. On the four datasets examined, the undetected proportions for children surpass those for adults by an average of 20.14\%. However, the overall performance of these pedestrian detectors in detecting males and females and dark-skin and light-skin groups does not exhibit a large difference, with only a 1.19\% and 0.44\% gap in undetected proportions. \textbf{(2)}~The studied pedestrian detectors reveal significant gender biases during night time, with a higher proportion of females going undetected compared to males. This situation may aggravate existing societal concerns about female safety during nighttime outings.
\textbf{(3)} In contrast to the commonly accepted fairness-performance trade-off, our findings suggest that pedestrian detectors can achieve enhanced fairness and detection performance under specific driving scenarios, such as those with higher brightness levels. 

To summarise, we make the following contributions:
\begin{itemize}[leftmargin=*]
\item We conduct the first comprehensive study on fairness testing of autonomous driving systems across various datasets and demographic attributes, evaluating eight widely-studied DL-based pedestrian detectors and uncovering significant fairness issues.
\item We augment four real-world datasets with manually labeled demographic information, resulting in 8,311 images with 16,070 gender labels, 20,115 age labels, and 3,513 skin tone labels.
\item We publicly release the data, demographic labels, and code used in this study~\cite{githublink} to facilitate future research on fairness of autonomous driving systems. 
\end{itemize}

\section{Background and Related Work}\label{preliminaries}
This study resides at the intersection of two increasingly important SE topics: software fairness and autonomous driving testing. To provide the necessary context, we begin by reviewing the background knowledge and relevant prior research in these areas.

\subsection{Software Fairness} 
\label{sec:se4fairness}
Fairness has gained significant attention in the SE community since its initial exploration by SE researchers in 2008 \cite{reFinkelsteinHMRZ08}. There have been various definitions of fairness in the literature. In this paper, we focus on group fairness, a concept extensively studied in software fairness research~\cite{10.1145/3468264.3468536,9402057,10.1145/3468264.3468565,Chakraborty2021BiasIM,inproceedings,icseGoharBR23,sigsoftBiswasR20,tosemmengdi,sigsoftZhang022,sigsoftNguyenBR23,icseChenZSH24}. Notably, group fairness closely aligns with legal regulations on fairness \cite{Barocas2016BigDD}, such as the adherence to the four-fifths rule, a cornerstone of US anti-discrimination law \cite{biddle2005adverse, DBLP:conf/kdd/FeldmanFMSV15, DBLP:conf/innovations/DworkHPRZ12, DBLP:conf/aistats/ZafarVGG17, NEURIPS2019_373e4c5d}. Consequently, testing and prioritizing group fairness when building software has emerged as an essential ethical duty and requirement for software engineers \cite{Chakraborty2021BiasIM}.

In the context of group fairness, certain personal characteristics that require protection against unfairness during decision-making are called sensitive attributes, also known as protected attributes~\cite{tseZhangHML22, CorbettDavies2018TheMA,DBcorrabs220707068}. Well-recognized sensitive attributes include race, sex, age, pregnancy, familial status, disability status, and more~\cite{chen2022fairness,sigsoftWanWHGBL23}. These sensitive attributes typically partition a population into distinct groups: a privileged group and an unprivileged group~\cite{chen2022fairness}. Group fairness entails the equal treatment of these groups by the same machine learning model. However, in practice, members of unprivileged groups often experience systematic disadvantages, resulting from unfair machine learning models. For instance, in the context of a pedestrian detection task, if age is deemed a sensitive attribute, the predictive model may exhibit a bias favoring the adult group over the child group. In this scenario, the adult group is considered the privileged group, while the child group becomes the unprivileged one.

Recently, Chen et al.~\cite{chen2022fairness} have presented a comprehensive survey of fairness testing research and analyzed its trend. This survey points out that the majority of existing work revolves around tabular data \cite{10.1145/3368089.3409704,10.1145/3468264.3468536,icseGoharBR23,9402057,10.1145/3583561}. For example, Biswas and Rajan \cite{10.1145/3368089.3409704} evaluated fairness of machine learning models on a crowd sourced platform using tabular datasets covering tasks such as credit risk prediction, income prediction, marketing, and loan application. Similarly, Chen et al. \cite{10.1145/3583561} conducted an empirical study on the group fairness achieved by state-of-the-art bias mitigation methods across eight commonly used tabular data-driven decision tasks.
In contrast, our paper centers on fairness testing for pedestrian detection in autonomous driving systems. We specifically examine three sensitive attributes (i.e., gender, age, and skin tone) that are recognizable in autonomous driving datasets. These sensitive attributes have been demonstrated to be the most widely considered ones in the fairness testing literature \cite{chen2022fairness}.

\subsection{Autonomous Driving Testing} 
Autonomous driving testing is a hot SE research topic, and researchers have proposed various testing techniques for autonomous driving systems~\cite{10.1145/3361566, 10.1145/3180155.3180220, 10.1145/3238147.3238187, 10.1145/3377811.3380422, DBLP:conf/issta/GuoF022,dacZuoLYLJ21}. 
For instance, Tian et al.~\cite{10.1145/3180155.3180220} proposed DeepTest, a novel technique using image transformations to emulate potential camera disturbances encountered in driving environments. Zhang et al. ~\cite{10.1145/3238147.3238187} presented DeepRoad, employing GANs to craft test images derived from actual driving scenarios. Zhou et al. ~\cite{10.1145/3377811.3380422}  introduced DeepBillboard, aiming to generate adversarial billboards that could lead to steering mistakes in autonomous vehicles. Guo et al.~\cite{DBLP:conf/issta/GuoF022} developed LiRTest, marking the first automated testing technique for LiDAR-equipped autonomous driving software.

While a substantial body of knowledge focuses on assessing the robustness and correctness properties of autonomous driving systems~\cite{tseZhangHML22}, to the best of our knowledge, only a few studies have explored the fairness properties, particularly in the pedestrian detection domain within autonomous driving, indicating that this area remains under-explored.

Pedestrian detection is a crucial process that identifies pedestrians within street-level images by providing their predicted locations along with corresponding bounding boxes and confidence scores ~\cite{Dollr2012PedestrianDA, 7780510}. Despite its significance, research on the fairness of pedestrian detection within autonomous driving is limited. 
Brandao~\cite{Brandao2019AgeAG} explored fairness in pedestrian detection, concentrating on classic machine learning-based methods. These classic techniques, reliant on manually-defined features, have been eclipsed by deep learning-based pedestrian detection approaches, now prevalent in the autonomous driving domain. Wilson et al. ~\cite{Wilson2019PredictiveII} focused on skin tone bias, confined their analysis to a single dataset and two general object detectors. Similarly, Kogure et al. ~\cite{Kogure2022AgeSN} explored age bias using a small-scale dataset and a detection method that is no longer state-of-the-art. 

In summary, current fairness studies in pedestrian detection suffer from a lack of variety in the pedestrian detectors evaluated, the datasets used, and the range of sensitive attributes explored. Furthermore, no previous work has explored how different environmental characteristics (e.g., brightness, contrast, and weather conditions) affect fairness.

To address this knowledge gap, our paper presents a comprehensive empirical study on revealing fairness issues in pedestrian detection. We conduct experiments using eight popular DL-based detection methods and four diverse testing datasets, encompassing different scenarios determined by a variety of factors such as brightness, contrast, and weather conditions. We focus on three widely considered sensitive attributes, i.e., gender, age, and skin tone.
The scale and diversity of our experiments enable us to provide comprehensive insights into the fairness circumstances of the existing pedestrian detectors.

\section{Experimental Design}
\label{approach}
This section introduces our research questions and experimental design.

\begin{figure}
\centering
\includegraphics[width=1\linewidth]{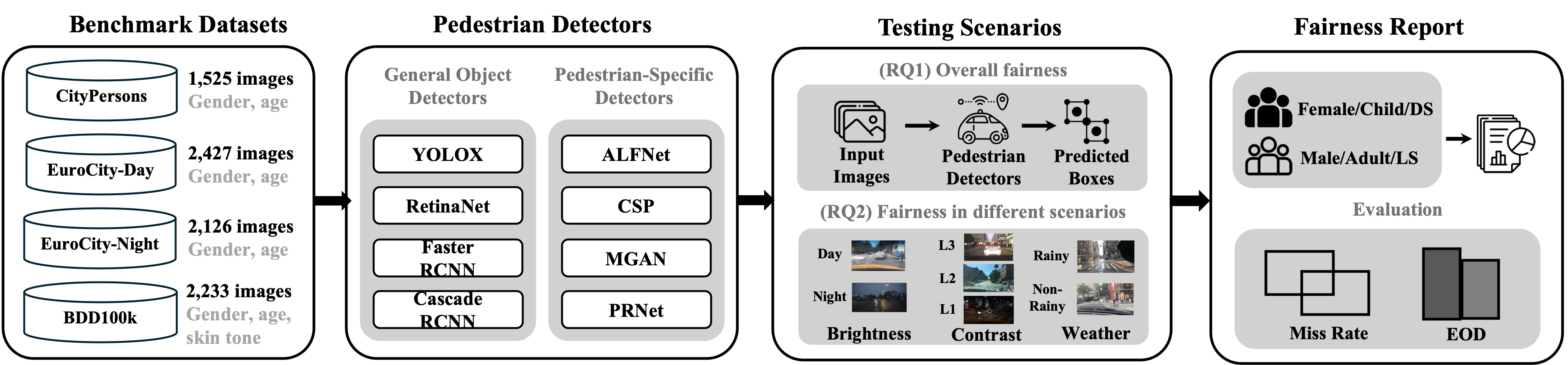}
\caption{Overview of our experimental settings.}
\label{fig1:Frame}
\end{figure}

\subsection{Research Questions}\label{RQ design}

We aim to answer the following research questions (RQs).

\noindent \textbf{RQ1 (Overall fairness):} \textit{To what extent do widely-studied pedestrian detectors exhibit unfairness concerning common sensitive attributes?} This RQ explores the performance difference of widely-studied pedestrian detectors when they are applied to different demographic groups, characterized by sensitive attributes including gender, age, and skin tone.

\noindent \textbf{RQ2 (Fairness in different scenarios):} \textit{What fairness do pedestrian detectors achieve in different brightness, contrast, and weather conditions?} We further investigate the fairness of commonly-studied pedestrian detectors in different autonomous driving scenarios by considering a variety of brightness (RQ2.1), contrast (RQ2.2), and weather conditions (RQ2.3) \cite{10.1145/3180155.3180220, 10.1145/3238147.3238187, 10.1145/3377811.3380422}. 

Figure \ref{fig1:Frame} illustrates our experimental settings to answer these RQs. In the following, we introduce the pedestrian detectors, benchmark datasets, evaluation metric, statistical analysis, and experimental details.

\subsection{Pedestrian Detectors}
In recent years, DL has revolutionized pedestrian detection approaches. We focus our analysis on eight DL-based pedestrian detectors that are widely studied in the autonomous driving community~\cite{10028728,Cao2020FromHT}. These detectors are pre-trained DL models that researchers and practitioners can directly use for pedestrian detection tasks. They can be classified into two categories: general object detectors and pedestrian-specific detectors~\cite{Hasan2020GeneralizablePD}. Next, we briefly introduce each category and the corresponding pedestrian detectors. 

Table \ref{table3:8 pedestrian detectors} provides an overview of these detectors. ``Detector'' shows the name of a pedestrian detector; 
``Backbone'' represents the pre-trained deep neural network used for extracting features from input images;
``Source'' indicates the framework/toolkit name and its source for a given pedestrian detector.

\begin{table}[tp]
\small
\caption{Pedestrian detectors.}
\label{table3:8 pedestrian detectors}
\centering
\begin{tabular}{c|lll}
\toprule
Type & Detector & Backbone & Source \\
\midrule
\multirowcell{4}{General object detectors} & YOLOX & - & \multirow{4}{*}{MMDetection \cite{MMDetectionModelZoo}} \\
 & RetinaNet & X-101-64x4d-FPN & \\
 & Faster RCNN & X-101-64x4d-FPN & \\
 & Cascade RCNN & X-101-64x4d-FPN & \\
\midrule
\multirowcell{4}{Pedestrian-specific detectors} & ALFNet & ResNet50 & ALFNet \cite{ALFNet}\\
 & CSP & ResNet50 & \multirow{2}{*}{Pedestron \cite{Pedestron}} \\
 & MGAN & VGG16 & \\
 & PRNet & ResNet50 & PRNet \cite{PRNet} \\
\bottomrule
\end{tabular}
\end{table}

\noindent\textbf{General object detectors}: General detectors can detect various objects such as cars, traffic lights, and pedestrians. They have great generalization ability but lack pedestrian-specific adaptation~\cite{Hasan2020GeneralizablePD}. They can be categorized into two categories: two-stage and one-stage detectors~\cite{10028728}. Two-stage detectors propose regions before feature extraction and classification, achieving high accuracy but slower speed; one-stage detectors complete all operations in one step, providing faster speed but lower accuracy. Both types involve trade-offs and are widely used for pedestrian detection. Hence, this paper selects detectors from both categories. For one-stage detectors, we adopt the widely-studied \textbf{YOLOX}~\cite{ge2021yolox} (a faster extension of the YOLO series~\cite{redmon2016you} used in Apollo's autonomous driving systems~\cite{Apollo}) and \textbf{RetinaNet}~\cite{lin2017focal} (which addresses the class imbalance problem). For two-stage detectors, we employ the \textbf{Faster RCNN}~\cite{NIPS2015_14bfa6bb} (one of the pioneering detectors in the RCNN family) and \textbf{Cascade RCNN}~\cite{cai2018cascade} (which achieves higher accuracy through a cascade of multiple CNNs to refine region proposals). These detectors have been extensively used in the autonomous driving literature~\cite{Hasan2020GeneralizablePD, Cao2020FromHT, Yu_2020_CVPR, 8634919}.

\noindent\textbf{Pedestrian-specific detectors}: 
Pedestrian-specific detectors use additional pedestrian-related information to improve detection performance~\cite{Hasan2020GeneralizablePD}. In this study, we investigate state-of-the-art pedestrian-specific detectors, including ALFNet~\cite{liu2018learning}, CSP~\cite{liu2019high}, MGAN \cite{pang2019mask}, and PRNet~\cite{song2020progressive}. \textbf{ALFNet} uses progressive detection heads on SSD~\cite{liu2016ssd} to refine initial anchors for improved detection accuracy. \textbf{CSP} introduces an anchor-free approach by locating center points and scaling pedestrians. \textbf{MGAN} uses visible-area bounding-box information to guide attention mask generation for occluded pedestrian detection. \textbf{PRNet} presents a novel progressive refinement network for occluded pedestrian detection.

\subsection{Benchmark Datasets} \label{benchmark datasets}

\subsubsection{Dataset Selection}\label{dataselection}
We perform our experiments on four real-world datasets that have been extensively studied by researchers to evaluate the performance of pedestrian detectors in autonomous driving~\cite{Cao2020FromHT, Hasan2020GeneralizablePD}.  These datasets consist of street-level images captured by cameras mounted on autonomous vehicles, showcasing pedestrians in diverse poses, sizes, and occlusion scenarios. Table~\ref{table1:Benchmark datasets} presents details about these datasets, including the sensitive attributes, the number of images in each dataset, and the respective time of day when these images were captured. Next, we briefly introduce each dataset:

\begin{table}[tp]
\small
\caption{Benchmark datasets.}
\label{table1:Benchmark datasets}
\centering
\begin{tabular}{l|lrl}
\toprule
Name & Sensitive Attributes & \#Images & Time \\
\midrule
CityPersons & gender, age & 1,525 & day \\
EuroCity-Day & gender, age & 2,427 & day \\
EuroCity-Night & gender, age & 2,126 & night\\
BDD100k & gender, age, skin tone & 2,233 & day, night\\
\bottomrule
\end{tabular}
\end{table}
\begin{itemize}[leftmargin=*]
\item\textbf{CityPersons} dataset~\cite{8099957} stands as the most widely-studied benchmark for evaluating pedestrian detectors~\cite{Cao2020FromHT, Hasan2020GeneralizablePD, liu2018learning, song2020progressive, liu2019high, pang2019mask}. Its test set includes 1,525 images captured across six cities, showcasing diverse weather conditions and street scenes.

\item\textbf{EuroCityPersons} dataset~\cite{8634919} contains 4,553 images gathered from seven European cities, encompassing both day and night time captures. The dataset can be categorized into two sets: 2,427 images captured during the day, and 2,126 images captured at night, referred to as the \textbf{EuroCity-Day} dataset and the \textbf{EuroCity-Night} dataset, respectively.
 
\item\textbf{Berkeley Driving} dataset (a.k.a., \textbf{BDD100k} dataset)~\cite{Yu_2020_CVPR} is an extensive driving dataset, including 2,233 images from 40 classes that are typical of driving scenes. These images were captured from four different cities, providing various times of the day. Notably, this dataset showcases a greater diversity of pedestrians than the CityPersons and EuroCityPersons datasets, including individuals with varied skin tones. 
\end{itemize}

\subsubsection{Sensitive Attribute Labeling} \label{sensitive attribute labeling}
We focus on three sensitive attributes: gender, age, and skin tone. These attributes are identifiable in autonomous driving images and are recognized as the three most extensively studied sensitive attributes in fairness testing literature \cite{chen2022fairness}.
To enable fairness analysis, we need datasets with labels that indicate these sensitive attributes of the humans depicted in the images. 
Among the datasets investigated herein, the only sensitive attribute already labeled is the skin tone (i.e., light-skin tone and dark-skin tone) for the BDD100k dataset.

We manually label gender and age for each of the datasets considered in this study. For skin tone, we only use the BDD100k dataset. As described in Section \ref{dataselection}, this dataset shows a greater diversity in skin tones compared to other datasets. This diversity is linked to the geographic locations where the datasets were collected. Specifically, BDD100k, collected in the United States, exhibits more varied skin tones. In contrast, other datasets from European cities, such as CityPersons collected in Germany, have significantly fewer dark-skin individuals. As reported~\cite{pewresearch2024facts, wikiAfroGermans, wikiDemographicsGermany}, there is only around 1\% representation of individuals with dark skin in Germany. During our labeling process, we indeed encountered scarcity of dark-skin individuals in the CityPersons, EuroCity-Day, and EuroCity-Night datasets, leading to our decision not to label these three datasets with skin tone attributes.

The labeling process involves two annotators to minimize the influence of labeling bias. We focus on images that align with the widely-adopted ``reasonable subset'' principle~\cite{Dollr2012PedestrianDA}, meaning that we label images containing labeled pedestrians with a height of at least 50 pixels and little to no occlusion. For such images, human annotators can label the sensitive attributes with high confidence~\cite{Brandao2019AgeAG}. Using the filtered datasets, the two annotators independently label the gender and age attributes for each image. For gender, we follow previous studies~\cite{Brandao2019AgeAG} and consider only two labels: male and female. As for age, in line with the literature~\cite{Brandao2019AgeAG, Kogure2022AgeSN}, we classify pedestrians into two labels: child and adult, based on their physical characteristics depicted in the images.

To ensure the reliability of the labeling procedure, both annotators independently label the gender and age attributes for each image. We use Cohen's Kappa ($\kappa$) \cite{Cohen1960ACO}, a widely-adopted metric for measuring inter-rater agreement, during the independent labeling process~\cite{sigsoftWenCLL00JL21, icseChenYLCLWL21,ChenCLW0L20,WangCZ23,LiuGCWZMWJ23,ZhangCWCLMMHL24}. The obtained $\kappa$ values for gender and age attributes in each of the four datasets are summarized in Table \ref{table2:kappa value}. According to the literature~\cite{Landis1977TheMO}, a $\kappa$ value between 0.81 and 1 signifies almost perfect agreement, as shown in Table \ref{table2:agreement level}, while a value between 0.61 and 0.8 indicates substantial agreement.
In our labeling process, we achieve substantial agreement in gender labeling for the EuroCity-Day dataset and almost perfect agreement for all other tasks. This high level of agreement underscores the reliability of our labeling procedure~\cite{sigsoftWenCLL00JL21, icseChenYLCLWL21,ChenCLW0L20,WangCZ23,LiuGCWZMWJ23,ZhangCWCLMMHL24}. In cases where the two annotators encounter conflicts, an arbitrator is involved in the discussion to reach a consensus. 
After the labeling process, the summary of the number of labeled pedestrian instances for each dataset is presented in Table \ref{table3:labeled instances}.

\begin{table}[tp]
\small
\caption{Cohen's $\kappa$ values for labeling gender and age.}
\label{table2:kappa value}
\centering
\begin{tabular}{lrr}
   \toprule
   Dataset & $\kappa_{gender}$  & $\kappa_{age}$  \\
   \midrule
   CityPersons  & 0.814 & 0.847 \\
   EuroCity-Day  & 0.800& 0.925 \\
   EuroCity-Night  & 0.870 & 0.847\\
   BDD100k & 0.854  & 0.828  \\
   \bottomrule
\end{tabular}
\end{table}

\begingroup
\captionsetup[table]{labelformat=simple, labelsep=colon}
\renewcommand{\tablename}{Table}
\begin{table}[tp]
\small
\caption{Cohen's $\kappa$ values and corresponding agreement levels of inter-rater agreement.}
\label{table2:agreement level}
\centering
\begin{tabular}{lrr}
   \toprule
   $\kappa$ values & Agreement level\\
   \midrule
   $<$ 0 & No agreement \\
   0.01-0.20 & Slight agreement \\
   0.21-0.40 & Fair agreement \\
   0.41-0.60 & Moderate agreement \\
   0.61-0.80 & Substantial agreement \\
   0.81-1 & Almost perfect agreement \\
   \bottomrule
\end{tabular}
\end{table}
\endgroup

\begin{table}[tp]
\small
\caption{Number of labeled pedestrian instances per dataset.}
\label{table3:labeled instances}
\centering
\begin{tabular}{lcccccc}
\toprule
& \multicolumn{2}{c}{Gender} & \multicolumn{2}{c}{Age} & \multicolumn{2}{c}{Skin tone} \\
\cmidrule(lr){2-3} \cmidrule(lr){4-5} \cmidrule(lr){6-7}
Dataset & Male & Female & Adult & Child & Light-skin & Dark-skin \\
\midrule
CityPersons & 2,357 & 1,822 & 4,568 & 233 & - & - \\
EuroCity-Day & 1,726 & 1,646 & 4,498 & 100 & - & - \\
EuroCity-Night & 1,265 & 1,318 & 4,165 & 68 & - & - \\
BDD100k & 3,457 & 2,479 & 6,293 & 190 & 2,724 & 789 \\
\midrule
Overall & \textbf{8,805} & \textbf{7,265} & \textbf{19,524} & \textbf{591} & \textbf{2,724} & \textbf{789} \\
\bottomrule
\end{tabular}
\end{table}

\subsubsection{Scenario Labeling} \label{scenario}
To deeply explore the fairness of pedestrian detectors across various driving scenarios (i.e., different brightness, contrast and weather conditions), we also need images labeled with scenario information. We therefore classify the images containing labeled pedestrians (5,933 out of the total 8,311 images in all datasets) into different scenarios, to enable our analysis in RQ2. An alternative approach to achieving this purpose is via generating images with different scenarios using existing test generation techniques from the autonomous driving literature. We do not choose this approach because generated images are not real images and can 
suffer from unnaturalness \cite{tseZhangHML22}.

\noindent \textbf{Labeling brightness.} Brightness represents the overall lightness or darkness of the image. To distinguish the brightness of the images, we use the time-of-day labels provided in the dataset annotations. Specifically, we categorize the images into ``day time'' and ``night time''. As shown in Table~\ref{table1:Benchmark datasets}, the CityPersons and EuroCity-Day datasets consist entirely of day time images, while the EuroCity-Night dataset consists entirely of night time images. These datasets are straightforward to categorize based on time of day, though they do not provide more granular brightness labels. For the BDD100k dataset, we use the detailed ``timeofday'' labels provided by Wilson et al. \cite{Wilson2019PredictiveII}, which include separate annotations for ``day time'' (covering dawn, dusk, and full daylight) and ``night time''. We apply these labels directly in our study to distinguish between brightness conditions.

\noindent \textbf{Labeling contrast.}
Contrast is the difference in brightness between objects in an image. To quantify the contrast of each image, we use the Root Mean Square (RMS) contrast measurement~\cite{Peli:90}, a standard measure in the computer vision literature~\cite{DBLP:journals/iajit/BhuiyanK18, DBLP:journals/ijon/JakhetiyaLJGG18, PELLI201310}. To apply the RMS measurement, we first need to convert all images into the gray-scale mode~\cite{Peli:90}. Then, we calculate the RMS contrast value for each image based on the converted version. A higher RMS contrast value indicates a greater contrast. To classify the images into different contrast levels, we identify the maximum RMS contrast value (which is 89.45) and the minimum RMS contrast value (which is 11.42) among all images. Then, we evenly divide this RMS contrast values range into three classes (each level can have sufficient images for statistical analysis), each covering an interval of 26.01 units (calculated as (max-min)/3). Each class represents a contrast level, labeled from level 1 to level 3, with higher levels indicating images with higher contrast. Then, we categorize the images into their respective contrast levels based on their RMS contrast values.

\noindent \textbf{Labeling weather conditions.} Common weather conditions studied in the autonomous driving literature include rain, fog, and snow~\cite{10.1145/3180155.3180220, 10.1145/3377811.3380422, 10.1145/3238147.3238187, 10.1145/3361566}. However, our datasets rarely contain images depicting fog and snow. This is due to the fact that our datasets are collected from real-world scenarios where fog and snow are infrequently encountered. The limited samples of snowy or foggy weather pose challenges for statistical analysis. As a result, we focus on rain as the weather condition of interest. Two annotators independently classify images containing labeled pedestrians into two categories: rainy and non-rainy. During this process, 1,856 images are not annotated because neither the two annotators nor the arbitrator could accurately distinguish the weather conditions.

To measure inter-rater agreement during manual labeling, we also use Cohen's Kappa ($\kappa$). The $\kappa$ value is 0.813, indicating almost perfect agreement~\cite{Landis1977TheMO}. This high level of agreement confirms the reliability of our labeling procedure. After scenario labeling, the summary of the number of images under different brightness, contrast, and weather conditions is presented in Table~\ref{table5:number of images in different classes}.

\begin{table*}[!tp]
  \centering
  \small
  \caption{Number of images in different brightness conditions, contrast levels, and weather conditions.}
    \begin{tabular}{cc|ccc|cc}
    \toprule
    \multicolumn{2}{c|}{Brightness conditions}                & \multicolumn{3}{c|}{Contrast levels}  & \multicolumn{2}{c}{Weather conditions} \\
    \midrule
    Day time & Night time & Level 1 & Level 2 & Level 3 & Rainy & Non-Rainy \\
    \midrule
    4,409  & 1,524 & 1,163  & 3,933  & 837  & 277  & 3,800 \\
    \bottomrule
    \end{tabular}
  \label{table5:number of images in different classes}
\end{table*}

\subsection{Evaluation Metric} \label{evalution}
There have been well-established quantitative measures for group fairness in the literature. The most widely-adopted fairness measures include SPD (Statistical Parity Difference), EOD (Equal Opportunity Difference), and AOD (Average Odds Difference)~\cite{sigsoftChenZSH22, chen2022fairness, 9402057}. Let a sensitive attribute be $A$, with 0 as the unprivileged group and 1 the privileged group; let the real classification label be $Y$ and the predicted label $\hat{Y}$, with 0 as the unfavorable class and 1 as the favorable class. In addition, we use $Pr$ to denote probability.

SPD quantifies the difference in the  probabilities of favorable outcomes between unprivileged and privileged groups:
\begin{equation}
SPD = Pr[\hat{Y} = 1|A = 0] - Pr[\hat{Y} = 1|A = 1]
\end{equation}

EOD quantifies the difference in the true-positive rates between unprivileged and privileged groups: 
\begin{equation}
EOD = Pr[\hat{Y} = 1|A = 0, Y = 1] - Pr[\hat{Y} = 1|A = 1, Y = 1]
\end{equation}

AOD quantifies the average difference between the false-positive rates and true-positive rates for unprivileged and privileged groups:
\begin{equation}
\begin{aligned}
AOD = \frac{1}{2}(&|Pr[\hat{Y} = 1|A = 0, Y = 0] - Pr[\hat{Y} = 1|A = 1, Y = 0]| \\
    &+ |Pr[\hat{Y} = 1|A = 0, Y = 1] - Pr[\hat{Y} = 1|A = 1, Y = 1]|)
\end{aligned}
\end{equation}

In the context of pedestrian detection, both SPD and EOD measure the disparity in proportions of successfully detected pedestrians between privileged and unprivileged groups. They also both express the difference in miss rates between privileged and unprivileged groups. 
\textbf{Miss Rate (MR)} is the most commonly-studied performance metric in pedestrian detection~\cite{Brandao2019AgeAG,Kogure2022AgeSN}, which quantifies the proportion of undetected pedestrians.
Formally, it is calculated as follows:
\begin{equation}
MR = 1 - \frac{TP}{TP+FN},
\end{equation}

where TP (true positive) refers to the number of successfully detected ground-truth bounding boxes, and FN (false negative) denotes the number of undetected ground-truth bounding boxes. Pedestrian detectors generate bounding box locations and confidence scores for recognized ``person'' instances in images. To assess whether a given ground-truth bounding box is successfully detected, the standard method in the literature is to use the Intersection over Union (IoU) metric~\cite{Dollr2012PedestrianDA}. The IoU metric quantifies the degree of overlap between the ground-truth bounding box and the detected bounding box. If the IoU value is greater than 50\%, the ground-truth bounding box is considered successfully detected \cite{Dollr2012PedestrianDA}. Otherwise, it is classified as undetected.

The calculation of AOD requires precise false-positive information, referring to instances where members of the negative class (non-pedestrians) are incorrectly classified as the positive class (pedestrians). As described in Section \ref{sensitive attribute labeling}, we adhere to the standard practice in the literature, where we focus on a ``reasonable subset'' of pedestrians within the images. This approach presents challenges in calculating precise false-positives for each group because the negative class (non-pedestrian) may also include instances that belong to the positive class (pedestrian). Therefore, we do not consider AOD in this study.

In summary, we use both SPD and EOD as fairness measures for evaluating pedestrian detectors. Since these two measures yield identical values for pedestrian detection, for the remainder of the paper, we present only \textbf{EOD}.

\subsection{Statistical Analysis}\label{stastana}
To assess the extent to which any observed unfairness is statistically significant (i.e., whether there is a significant difference in the miss rate between privileged and unprivileged groups), we use the two-proportion z-test \cite{casella2007statistical}. This statistical test is widely used to analyze differences between proportions~\cite{sigsoftNiuWG16, VeizagaATSB21}. A result is deemed significant only if the obtained $p$-value falls below a predetermined threshold (in our case, 0.05, a widely-accepted
threshold in the fairness literature~\cite{sigsoftChenZSH22,10.1145/3583561}). For instance, in evaluating whether there exists a difference between the miss rates for male ($MR_{male}$) and female ($MR_{female}$) individuals detected by a pedestrian detector, the null hypothesis assumes that $MR_{male}$ is equal to $MR_{female}$. If the resulting $p$-value is lower than 0.05, we reject the null hypothesis, indicating a significant difference between $MR_{male}$ and $MR_{female}$.

\subsection{Experimental Details}\label{details}
The experiments are implemented based on open-source frameworks of each pedestrian detector. For general object detectors, we select the pre-trained models with the highest accuracy from the  the MMdetection model zoo ~\cite{MMDetectionModelZoo}. For pedestrian-specific detectors, we employ pre-trained models available from their respective public repositories ~\cite{Pedestron, PRNet, ALFNet}.

To ensure the reliability of our results, all experiments are repeated ten times. The final results are derived by calculating the average across these ten iterations.

All experiments are performed on a platform equipped with 64GB RAM, 2.5GHz Intel Xeon (R) v3 Dual CPUs, and one NVIDIA GeForce RTX 2080 Ti GPU. YOLOX, RetinaNet, Faster RCNN, and Cascade RCNN are implemented using PyTorch 1.8.1 and Python 3.7 on Ubuntu 18.04 LTS, following the MMdetection configuration \cite{mmdetection}. CSP and MGAN use PyTorch 1.10.0 and Python 3.8 on Ubuntu 18.04 LTS, adhering to the Pedestron configuration \cite{hasan2022pedestrian}. ALFNet \cite{liu2018learning} and PRNet~\cite{song2020progressive} are implemented using Keras 2.0.6, Tensorflow 1.4.0, and Python 2.7 on Ubuntu 16.04 LTS.

\section{Results}\label{results}
This section answers our RQs based on experimental results.

\subsection{RQ1: Overall Fairness}
RQ1 investigates the overall fairness of eight state-of-the-art pedestrian detectors regarding gender, age, and skin tone. First, for each detector, we compute the miss rate (MR) for different demographic groups and calculate EOD based on the MR results over all the datasets that we study. We also use the two-proportion z-test to determine the significance of any observed unfairness, as described in Section \ref{stastana}. Table \ref{table6:overall-performance of age, gender and skin} presents the results, with significant unfairness results highlighted in shading. In the following, we analyze the results for gender, age, and skin tone, respectively.

\begin{table*}[htbp]
  \setlength{\tabcolsep}{2pt}
  \centering
  \small
  \caption{(RQ1) Overall fairness in pedestrian detection across gender, age, and skin tone. Statistically significant biases, indicated by EOD (i.e., miss rate difference), are shaded. On average, detectors display comparable miss rates for female and male pedestrians, as well as for dark-skin and light-skin individuals. However, concerning age, significant bias is observed, as pedestrian detectors exhibit a 20.14\% higher miss rate for children compared to adults.}
    \scalebox{0.73}{
    \begin{tabular}{c|ccc|ccc|ccc}
    \toprule
    \textbf{Detectors} & \textbf{MR Male} & \textbf{MR Female} & \textbf{EOD (Gender)} & \textbf{MR Adult} & \textbf{MR Child} & \textbf{EOD (Age)} & \textbf{MR Light-skin} & \textbf{MR Dark-skin} & {\textbf{EOD (Skin)}} \\
    \midrule
    YOLOX & 9.78\% & 10.50\% & -0.72\% & \colorcell12.64\% & \colorcell42.47\% & \colorcell-29.83\% & 5.21\% & 3.80\% & 1.41\% \\
    RetinaNet & \colorcell10.72\% & \colorcell12.59\% & \colorcell-1.87\% & \colorcell14.36\% & \colorcell44.33\% & \colorcell-29.97\% & \colorcell8.33\% & \colorcell4.44\% & \colorcell3.90\% \\
    Faster RCNN & 3.80\% & 4.13\% & -0.32\% & \colorcell5.24\% & \colorcell26.06\% & \colorcell-20.82\% & \colorcell5.91\% & \colorcell3.30\% & \colorcell2.62\% \\
    Cascade RCNN & 3.87\% & 4.16\% & -0.28\% & \colorcell5.12\% & \colorcell26.57\% & \colorcell-21.44\% & \colorcell5.21\% & \colorcell3.04\% & \colorcell2.17\% \\
    ALFNet & \colorcell30.86\% & \colorcell32.90\% & \colorcell-2.04\% & \colorcell36.62\% & \colorcell53.47\% & \colorcell-16.85\% & 42.55\% & 43.98\% & -1.43\% \\
    CSP   & \colorcell33.49\% & \colorcell35.25\% & \colorcell-1.76\% & \colorcell37.74\% & \colorcell50.42\% & \colorcell-12.68\% & 61.38\% & 64.77\% & -3.39\% \\
    MGAN  & 29.76\% & 30.97\% & -1.21\% & \colorcell33.58\% & \colorcell46.53\% & \colorcell-12.95\% & 52.94\% & 54.88\% & -1.94\% \\
    PRNet & 40.28\% & 41.62\% & -1.34\% & \colorcell44.69\% & \colorcell61.25\% & \colorcell-16.56\% & 59.65\% & 59.44\% & 0.21\% \\
    \midrule
    {\textbf{Average }} & {\textbf{20.32\%}} & {\textbf{21.52\%}} & {\textbf{-1.19\%}} & \colorcell{\textbf{23.75\%}} & \colorcell{\textbf{43.89\%}} & \colorcell{\textbf{-20.14\%}} & {\textbf{30.15\%}} & {\textbf{29.71\%}} & {\textbf{0.44\%}} \\
    \bottomrule
\end{tabular}}
  \label{table6:overall-performance of age, gender and skin}
\end{table*}
\noindent \textbf{Gender.} As shown in Table \ref{table6:overall-performance of age, gender and skin}, on average, the miss rate difference between female and male pedestrians is merely 1.19\% ($p$-value $\textgreater$ 0.05), indicating that this difference is not statistically significant.

\begin{figure*}
\centering
\includegraphics[width=1\linewidth]{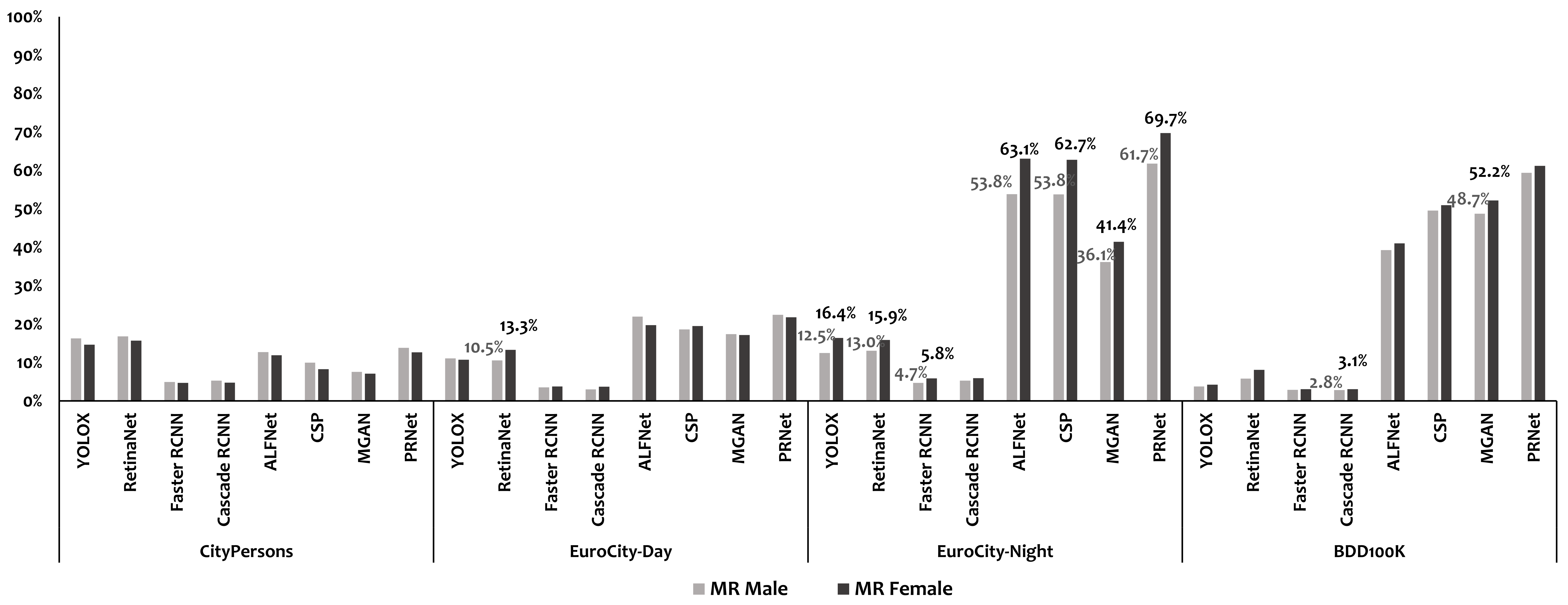}
\caption{(RQ1) Miss rates of pedestrian detectors for females and males across datasets. Statistically significant gender biases are indicated by labeled miss rate values. In CityPersons and EuroCity-Day datasets with only day time data, only one detector in the EuroCity-Day dataset exhibits significant gender bias. However, in the EuroCity-Night dataset, seven out of eight detectors show significantly higher miss rates for females, revealing bias in female detection.}
\label{fig4:MR_gender_4datasets}
\end{figure*}

Furthermore, we analyze the miss rate difference achieved by each pedestrian detector across the four datasets used in our study. Figure \ref{fig4:MR_gender_4datasets} illustrates the results. Significant gender biases (i.e., significant miss rate differences) are indicated by labeled miss rate values.
We observe that in the CityPersons and EuroCity-Day datasets, containing only day time data, only one detector in the EuroCity-Day dataset exhibits significant differences in miss rates between females and males. For the remaining results, there are no notable differences in miss rates between genders. However, results from the EuroCity-Night dataset present a contrasting observation, where seven of eight detectors exhibit a significantly higher miss rate for females, indicating bias in detecting females. In the BDD100k dataset that includes both day time and night time images, the miss rate difference is less pronounced. These observations motivate us to hypothesize that brightness conditions may influence the fairness of pedestrian detectors, which is further investigated in RQ2.

\noindent \textbf{Age.}
As observed in Table \ref{table6:overall-performance of age, gender and skin}, pedestrian detection exhibits large age bias, with all studied detectors demonstrating significantly higher miss rates for children compared to adults ($p$-value $\textless$ 0.05). On average, the miss rate difference between children and adults is 20.14\%. 

\begin{figure*}
\centering
\includegraphics[width=1\linewidth]{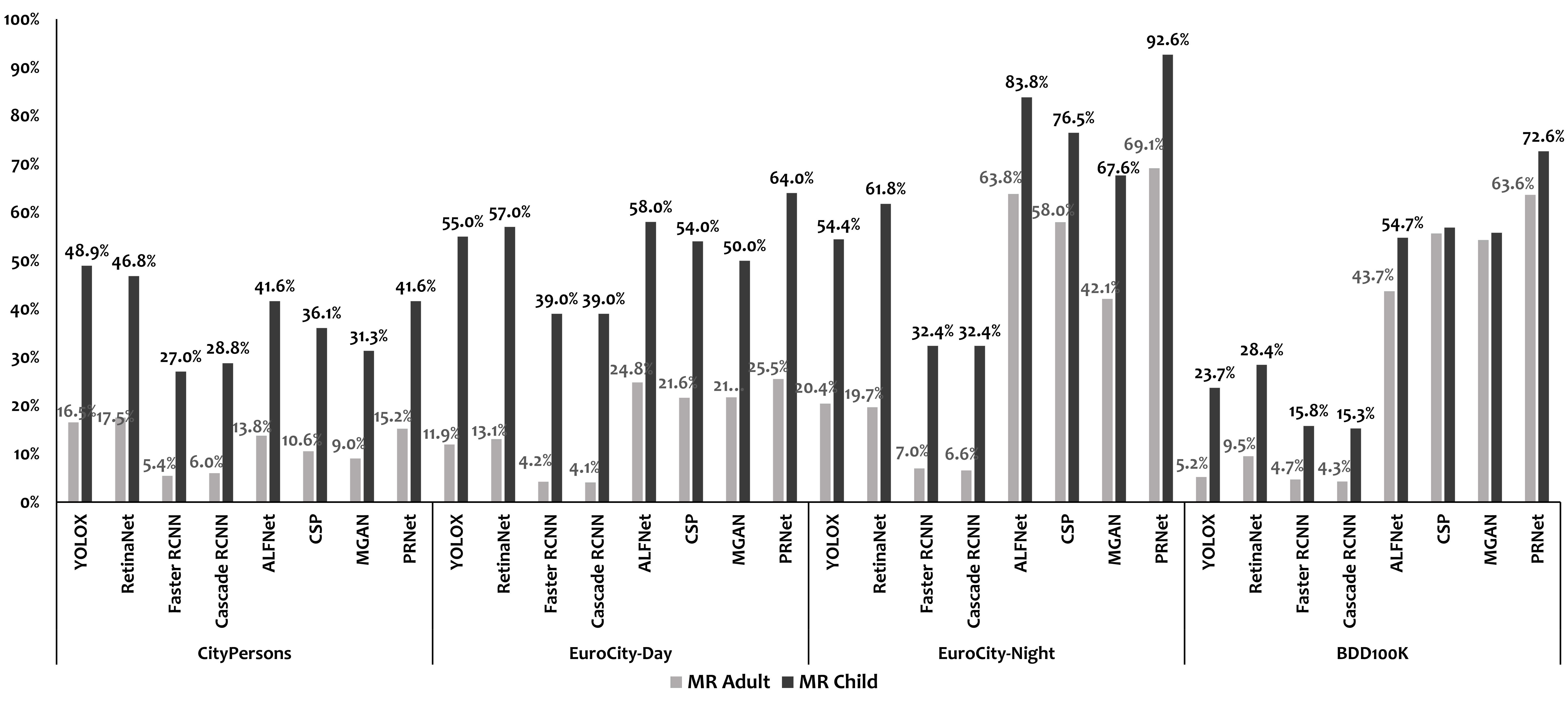}
\caption{(RQ1) Miss rates of pedestrian detectors for children and adults across datasets.  Statistically significant age biases are labeled with miss rate values. In 30 out of 32 scenarios (comprising four datasets and eight detectors), children have significantly higher miss rates than adults.}
\label{fig5:MR_age_4datasets}
\end{figure*}

Furthermore, we illustrate the miss rate difference across four datasets in Figure \ref{fig5:MR_age_4datasets}, with significant age biases indicated by labeled miss rate values. We observe that the miss rate of children is consistently higher than that of adults across all the datasets and all detectors. In particular, out of the total 32 results (combinations of four datasets and eight pedestrian detection models), 30 exhibit a statistically significant miss rate difference ($p$-value < 0.05). This indicates that the bias favoring adults is not specific to a particular dataset or detector, highlighting a strong unfairness between adults and children. 

\begin{figure}
\centering
\includegraphics[width=0.5\linewidth]{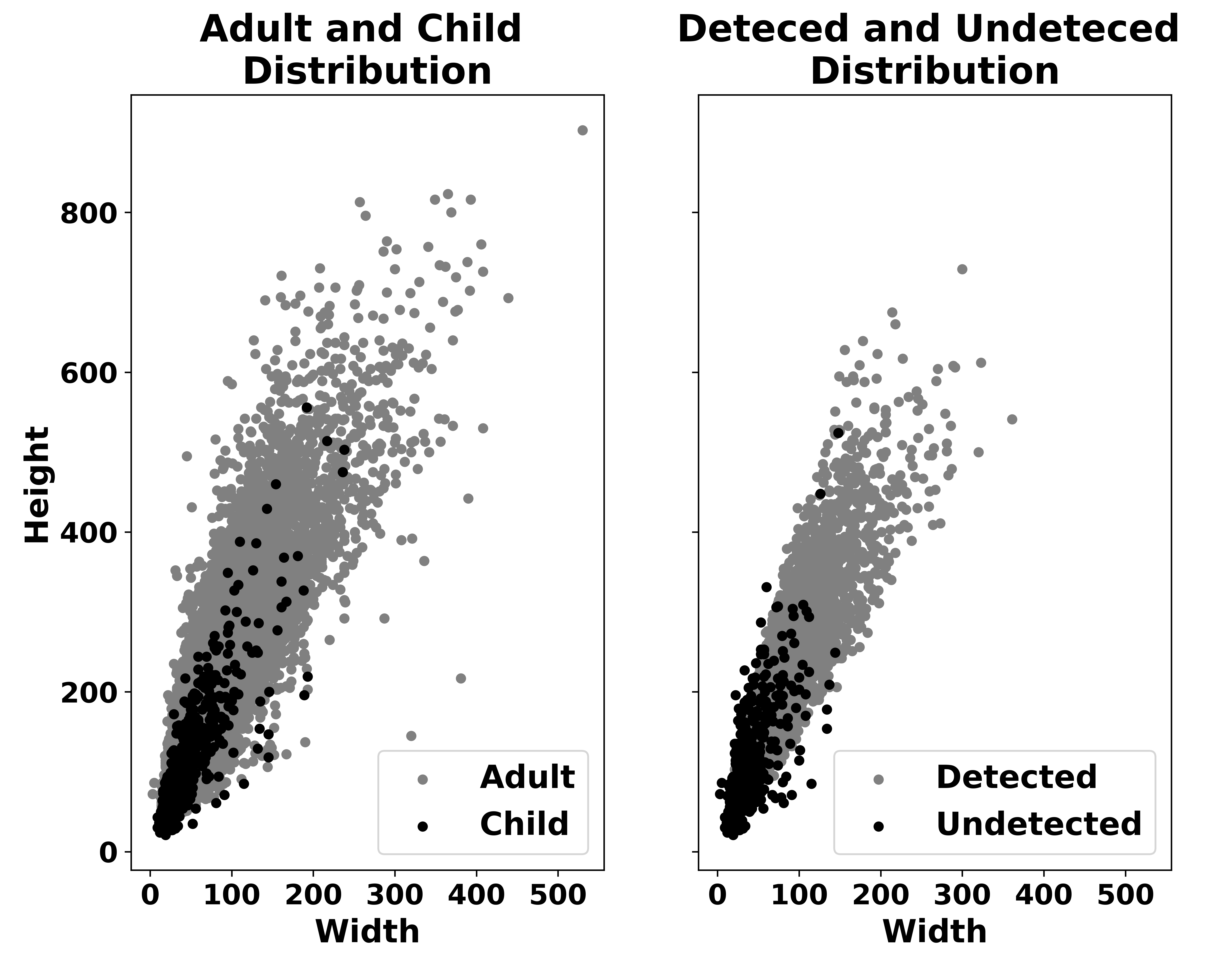}
\caption{(RQ1) Bounding box size distributions of adults and children (left) and bounding box size distributions of undetected and detected pedestrians (right). We observe that both children and undetected pedestrians tend to have smaller bounding boxes.} 
\label{fig5:distribution}
\end{figure}

The age bias may be attributed to the inherent challenge of detecting small objects, owing to the limited information provided by small bounding boxes \cite{7780510,5206631,Kogure2022AgeSN,Cao2020FromHT}. Given that children generally have smaller bodies compared to adults, their bounding boxes in the images also tend to be smaller. 
To demonstrate this, we analyze the distribution of bounding box sizes for pedestrians detected and undetected by all pedestrian detectors, as well as the distribution of ground-truth bounding box sizes for both adults and children. The results, presented in Figure \ref{fig5:distribution}, reveal a correspondence between these distributions. Specifically, the undetected bounding boxes and the ground-truth bounding boxes for children do not exceed 400 pixels in height and 200 pixels in width. This analysis shows that children, as well as undetected pedestrians, tend to have smaller bounding boxes.

\noindent \textbf{Skin tone.}
As displayed in Table \ref{table6:overall-performance of age, gender and skin}, current pedestrian detectors exhibit minimal bias between light-skin and dark-skin individuals, with an average miss rate difference of just 0.44\%. Specifically, among the four pedestrian-specific detectors, none show significant skin tone bias, whereas among the four general object detectors, three display significant skin tone bias, with miss rate differences ranging from 2.17\% to 3.9\%. These disparities are notably smaller than those observed in age bias.

\finding{\textbf{RQ1:} On average, the eight state-of-the-art pedestrian detectors that we study exhibit no significant performance difference across gender and skin tone, but notable bias across age. In particular, the detectors show a 20.14\% higher miss rate for children compared to adults.}

\subsection{RQ2: Fairness in Different Scenarios}
RQ2 evaluates the fairness of state-of-the-art pedestrian detectors under different real-world autonomous driving scenarios.

\subsubsection{RQ2.1: Different brightness conditions.} As introduced in Section \ref{scenario}, we consider two brightness conditions: day time and night time.

We first evaluate the overall miss rates of the eight pedestrian detectors under day time and night time. The results, presented in Table \ref{table7:day time and night time}, show a noticeable increase in average miss rates during night time compared to day time for each demographic group. For example, the average miss rates for males and females at night are 33.20\% and 36.57\%, respectively, compared to 17.24\% and 17.32\% during the day. A similar pattern is observed for age and skin tone. This indicates that the transition from day time to night time influences the performance of pedestrian detectors, with statistically significant higher miss rates observed at night.

Then, we investigate whether the performance is equally decreased for different demographic groups. Specifically, we explore the fairness change from day time to night time. Table \ref{table7:day time and night time} shows the results, with statistically significant biases (i.e., miss rate differences) emphasized in shading. 

Regarding gender, we observe a shift in the miss rate difference between males and females from day time to night time. During the day, there is only a slight -0.09\% difference for males and females. However, during night time, the difference increases to -3.37\% with statistically significant, indicating a notable change in fairness. In the night time condition, all pedestrian detectors exhibit higher miss rates for females compared to males, with six of them showing statistically significant differences. This potentially worsens concerns regarding female safety during nighttime out, a prevalent societal issue \cite{wang2023young}.

For age, all detectors exhibit significant biases during both day time and night time. Moreover, the miss rate difference for children and adults increases from day time to night time, with the average difference increasing from -22.45\% during the day to -25.48\% at night, which is statistically significant. This suggests a higher probability of children being undetected during night time. 

For skin tone, the miss rate difference between dark-skin and light-skin groups increases from day time to night time, with the average difference increasing from 0.15\% at day time to 3.16\% at night time. Nevertheless, the overall skin tone bias is not statistically significant during both day time and night time.

\begin{table*}[!tp]
  \setlength{\tabcolsep}{2pt}
  \centering
  \footnotesize
   \caption{(RQ2.1) Miss rates and EOD of each pedestrian detector under day time and night time. Statistically significant unfairness results are shaded. We find that reduced brightness conditions not only decrease the performance of pedestrian detectors but also exacerbate their biases. Notably, while pedestrian detectors generally do not exhibit significant gender bias during day time, biases against females become pronounced with six out of eight detectors showing significant biases during night time.}
\scalebox{0.88}{
    \begin{tabular}{c|ccc|ccc}
    \toprule
    \multicolumn{7}{c}{\textbf{Gender}} \\
    \midrule
    \multirow{2}[4]{*}{\textbf{Detectors}} & \multicolumn{3}{c|}{\textbf{Day time}} & \multicolumn{3}{c}{\textbf{Night time}} \\
\cmidrule{2-7}          & \textbf{MR Male} & \textbf{MR Female} & \textbf{EOD} & \textbf{MR Male} & \textbf{MR Female} & \textbf{EOD} \\
    \midrule
    YOLOX & 9.73\% & 9.38\% & 0.35\% & \colorcell9.99\% & \colorcell14.53\% & \colorcell-4.54\% \\
    RetinaNet & \colorcell10.42\% & \colorcell11.79\% & \colorcell-1.37\% & \colorcell11.99\% & \colorcell15.48\% & \colorcell-3.49\% \\
    Faster RCNN & 3.66\% & 3.59\% & 0.07\% & \colorcell4.41\% & \colorcell6.06\% & \colorcell-1.66\% \\
    Cascade RCNN & 3.58\% & 3.68\% & -0.10\% & 5.11\% & 5.87\% & -0.76\% \\
    ALFNet & 24.52\% & 24.39\% & 0.13\% & \colorcell57.29\% & \colorcell63.42\% & \colorcell-6.14\% \\
    CSP   & 26.76\% & 26.38\% & 0.38\% & \colorcell61.57\% & \colorcell67.09\% & \colorcell-5.51\% \\
    MGAN  & 25.48\% & 26.29\% & -0.81\% & 47.59\% & 47.76\% & -0.17\% \\
    PRNet & 33.73\% & 33.07\% & 0.66\% & \colorcell67.63\% & \colorcell72.33\% & \colorcell-4.70\% \\
    \midrule
    \textbf{Average} & \textbf{17.24\%} & \textbf{17.32\%} & \textbf{-0.09\%} & \colorcell\textbf{33.20\%} & \colorcell\textbf{36.57\%} & \colorcell\textbf{-3.37\%} \\
    \midrule
    \multicolumn{7}{c}{\textbf{Age}} \\
    \midrule
    \multirow{2}[4]{*}{\textbf{Detectors}} & \multicolumn{3}{c|}{\textbf{Day time}} & \multicolumn{3}{c}{\textbf{Night time}} \\
\cmidrule{2-7}          & \textbf{MR Adult} & \textbf{MR Child} & \textbf{EOD} & \textbf{MR Adult} & \textbf{MR Child} & \textbf{EOD} \\
    \midrule
    YOLOX & \colorcell10.80\% & \colorcell40.77\% & \colorcell-29.97\% & \colorcell17.81\% & \colorcell54.93\% & \colorcell-37.12\% \\
    RetinaNet & \colorcell12.73\% & \colorcell41.92\% & \colorcell-29.19\% & \colorcell18.97\% & \colorcell61.97\% & \colorcell-43.01\% \\
    Faster RCNN & \colorcell4.50\% & \colorcell25.00\% & \colorcell-20.50\% & \colorcell7.34\% & \colorcell33.80\% & \colorcell-26.46\% \\
    Cascade RCNN & \colorcell4.52\% & \colorcell25.58\% & \colorcell-21.05\% & \colorcell6.81\% & \colorcell33.80\% & \colorcell-26.99\% \\
    ALFNet & \colorcell26.52\% & \colorcell49.42\% & \colorcell-22.90\% & \colorcell65.10\% & \colorcell83.10\% & \colorcell-18.00\% \\
    CSP   & \colorcell28.64\% & \colorcell46.92\% & \colorcell-18.28\% & \colorcell63.42\% & \colorcell76.06\% & \colorcell-12.64\% \\
    MGAN  & \colorcell27.87\% & \colorcell43.46\% & \colorcell-15.59\% & \colorcell49.72\% & \colorcell69.01\% & \colorcell-19.30\% \\
    PRNet & \colorcell34.80\% & \colorcell56.92\% & \colorcell-22.13\% & \colorcell72.60\% & \colorcell92.96\% & \colorcell-20.36\% \\
    \midrule
    \textbf{Average} & \colorcell\textbf{18.80\%} & \colorcell\textbf{41.25\%} & \colorcell\textbf{-22.45\%} & \colorcell\textbf{37.72\%} & \colorcell\textbf{63.20\%} & \colorcell\textbf{-25.48\%} \\
    \midrule
    \multicolumn{7}{c}{\textbf{Skin Tone (LS: Light Skin, DS: Dark Skin)}} \\
    \midrule
    \multirow{2}[4]{*}{\textbf{Detectors}} & \multicolumn{3}{c|}{\textbf{Day time}} & \multicolumn{3}{c}{\textbf{Night time}} \\
\cmidrule{2-7}          & \textbf{MR LS} & \textbf{MR DS} & \textbf{EOD} & \textbf{MR LS} & \textbf{MR DS} & \textbf{EOD} \\
    \midrule
    YOLOX & 4.89\% & 4.03\% & 0.86\% & 7.06\% & 2.52\% & 4.53\% \\
    RetinaNet & \colorcell7.44\% & \colorcell4.33\% & \colorcell3.11\% & \colorcell13.38\% & \colorcell5.04\% & \colorcell8.34\% \\
    Faster RCNN & \colorcell5.36\% & \colorcell3.13\% & \colorcell2.23\% & 9.00\% & 4.20\% & 4.80\% \\
    Cascade RCNN & \colorcell4.84\% & \colorcell2.99\% & \colorcell1.86\% & 7.30\% & 3.36\% & 3.94\% \\
    ALFNet & 37.05\% & 38.66\% & -1.61\% & 73.48\% & 73.95\% & -0.47\% \\
    CSP   & 55.90\% & 59.85\% & -3.95\% & 92.21\% & 92.44\% & -0.22\% \\
    MGAN  & 47.12\% & 49.70\% & -2.58\% & 85.64\% & 84.03\% & 1.61\% \\
    PRNet & 54.39\% & 54.63\% & -0.24\% & 89.29\% & 86.55\% & 2.74\% \\
    \midrule
    \textbf{Average} & \textbf{27.31\%} & \textbf{27.16\%} & \textbf{0.15\%} & \textbf{47.17\%} & \textbf{44.01\%} & \textbf{3.16\%} \\
    \bottomrule
\end{tabular}}
  \label{table7:day time and night time}
\end{table*}

\finding{\textbf{RQ2.1: } Lower brightness conditions not only diminish the performance of pedestrian detectors but also exacerbate their bias. Particularly, during day time, pedestrian detectors generally do not exhibit significant gender bias, whereas six out of eight detectors demonstrate significant biases against females during night time.}

\subsubsection{RQ2.2: Different contrast levels.}

\begin{table*}[!tp]
  \setlength{\tabcolsep}{2pt}
  \centering
  \footnotesize
   \caption{(RQ2.2) Miss rates and EOD of each pedestrian detector under different contrast levels. On average, detectors exhibit the most biased results (i.e., the largest absolute value of EOD) under level 2, while demonstrating the fairest outcomes under level 3.}
\scalebox{0.88}{
    \begin{tabular}{c|ccc|ccc|ccc}
    \toprule
    \multicolumn{10}{c}{\textbf{Gender}} \\
    \midrule
    \multirow{2}[4]{*}{\textbf{Detectors}} & \multicolumn{3}{c|}{\textbf{level 3}} & \multicolumn{3}{c|}{\textbf{level 2}} & \multicolumn{3}{c}{\textbf{level 1}} \\
\cmidrule{2-10}          & \textbf{Male MR} & \textbf{Female MR} & \textbf{EOD} & \textbf{Male MR} & \textbf{Female MR} & \textbf{EOD} & \textbf{Male MR} & \textbf{Female MR} & \textbf{EOD} \\
    \midrule
    YOLOX & 6.30\% & 5.21\% & 1.09\% & 9.42\% & 10.42\% & -1.00\% & 13.99\% & 14.71\% & -0.72\% \\
    RetinaNet & 7.89\% & 9.46\% & -1.56\% & \colorcell10.47\% & \colorcell12.76\% & \colorcell-2.29\% & 13.99\% & 14.24\% & -0.25\% \\
    Faster RCNN & \colorcell3.83\% & \colorcell2.34\% & \colorcell1.49\% & 3.66\% & 4.16\% & -0.50\% & 4.34\% & 5.32\% & -0.98\% \\
    Cascade RCNN & 3.43\% & 2.44\% & 0.98\% & 3.65\% & 4.26\% & -0.62\% & 5.12\% & 5.01\% & 0.11\% \\
    ALFNet & 33.33\% & 31.67\% & 1.66\% & \colorcell30.28\% & \colorcell32.66\% & \colorcell-2.38\% & \colorcell31.09\% & \colorcell34.74\% & \colorcell-3.65\% \\
    CSP   & 40.11\% & 37.41\% & 2.70\% & \colorcell32.61\% & \colorcell34.80\% & \colorcell-2.19\% & \colorcell31.54\% & \colorcell35.45\% & \colorcell-3.90\% \\
    MGAN  & 38.52\% & 38.89\% & -0.38\% & \colorcell29.25\% & \colorcell31.23\% & \colorcell-1.98\% & 24.61\% & 24.10\% & 0.51\% \\
    PRNet & 48.41\% & 46.97\% & 1.43\% & \colorcell39.45\% & \colorcell41.52\% & \colorcell-2.06\% & 36.92\% & 38.11\% & -1.19\% \\
    \midrule
    \textbf{Average} & \textbf{22.73\%} & \textbf{21.80\%} & \textbf{0.93\%} & \colorcell\textbf{19.85\%} & \colorcell\textbf{21.48\%} & \colorcell\textbf{-1.63\%} & \textbf{20.20\%} & \textbf{21.46\%} & \textbf{-1.26\%} \\
    \midrule
    \multicolumn{10}{c}{\textbf{Age}} \\
    \midrule
    \multirow{2}[4]{*}{\textbf{Detectors}} & \multicolumn{3}{c|}{\textbf{level 3}} & \multicolumn{3}{c|}{\textbf{level 2}} & \multicolumn{3}{c}{\textbf{level 1}} \\
\cmidrule{2-10}          & \textbf{Adult MR} & \textbf{Child MR} & \textbf{EOD} & \textbf{Adult MR} & \textbf{Child MR} & \textbf{EOD} & \textbf{Adult MR} & \textbf{Child MR} & \textbf{EOD} \\
    \midrule
    YOLOX & \colorcell6.91\% & \colorcell32.89\% & \colorcell-25.98\% & \colorcell12.05\% & \colorcell40.92\% & \colorcell-28.87\% & \colorcell18.23\% & \colorcell53.23\% & \colorcell-35.00\% \\
    RetinaNet & \colorcell10.28\% & \colorcell36.84\% & \colorcell-26.56\% & \colorcell13.85\% & \colorcell41.94\% & \colorcell-28.10\% & \colorcell18.70\% & \colorcell56.45\% & \colorcell-37.75\% \\
    Faster RCNN & \colorcell4.00\% & \colorcell23.68\% & \colorcell-19.69\% & \colorcell5.08\% & \colorcell26.09\% & \colorcell-21.00\% & \colorcell6.56\% & \colorcell27.42\% & \colorcell-20.86\% \\
    Cascade RCNN & \colorcell3.33\% & \colorcell19.74\% & \colorcell-16.41\% & \colorcell5.05\% & \colorcell25.58\% & \colorcell-20.52\% & \colorcell6.48\% & \colorcell33.87\% & \colorcell-27.39\% \\
    ALFNet & \colorcell34.26\% & \colorcell52.63\% & \colorcell-18.37\% & \colorcell35.42\% & \colorcell54.73\% & \colorcell-19.32\% & 42.24\% & 50.00\% & -7.76\% \\
    CSP   & \colorcell40.72\% & \colorcell53.95\% & \colorcell-13.23\% & \colorcell36.92\% & \colorcell52.17\% & \colorcell-15.25\% & 38.72\% & 42.74\% & -4.02\% \\
    MGAN  & 40.38\% & 44.74\% & -4.35\% & \colorcell33.20\% & \colorcell48.85\% & \colorcell-15.65\% & \colorcell30.68\% & \colorcell40.32\% & \colorcell-9.65\% \\
    PRNet & \colorcell48.33\% & \colorcell68.42\% & \colorcell-20.09\% & \colorcell43.46\% & \colorcell63.17\% & \colorcell-19.71\% & 46.67\% & 50.81\% & -4.14\% \\
    \midrule
    \textbf{Average} & \colorcell\textbf{23.53\%} & \colorcell\textbf{41.61\%} & \colorcell\textbf{-18.08\%} & \colorcell\textbf{23.13\%} & \colorcell\textbf{44.18\%} & \colorcell\textbf{-21.05\%} & \colorcell\textbf{26.04\%} & \colorcell\textbf{44.35\%} & \colorcell\textbf{-18.32\%} \\
    \midrule
    \multicolumn{10}{c}{\textbf{Skin Tone (LS: Light Skin, DS: Dark Skin)}} \\
    \midrule
    \multirow{2}[4]{*}{\textbf{Detectors}} & \multicolumn{3}{c|}{\textbf{level 3}} & \multicolumn{3}{c|}{\textbf{level 2}} & \multicolumn{3}{c}{\textbf{level 1}} \\
\cmidrule{2-10}          & \textbf{LS MR} & \textbf{DS MR} & \textbf{EOD} & \textbf{LS MR} & \textbf{DS MR} & \textbf{EOD} & \textbf{LS MR} & \textbf{DS MR} & \textbf{EOD} \\
    \midrule
    YOLOX & 4.15\% & 4.91\% & -0.75\% & \colorcell5.75\% & \colorcell3.35\% & \colorcell2.40\% & 2.52\% & 2.17\% & 0.35\% \\
    RetinaNet & 7.27\% & 5.28\% & 1.99\% & \colorcell8.60\% & \colorcell4.18\% & \colorcell4.41\% & 10.08\% & 2.17\% & 7.91\% \\
    Faster RCNN & 5.04\% & 4.91\% & 0.14\% & \colorcell6.16\% & \colorcell2.51\% & \colorcell3.65\% & 6.72\% & 2.17\% & 4.55\% \\
    Cascade RCNN & 4.01\% & 4.53\% & -0.52\% & \colorcell5.64\% & \colorcell2.09\% & \colorcell3.55\% & 5.04\% & 4.35\% & 0.69\% \\
    ALFNet & 37.09\% & 36.60\% & 0.49\% & 42.62\% & 45.40\% & -2.78\% & 72.27\% & 71.74\% & 0.53\% \\
    CSP   & 53.26\% & 57.74\% & -4.47\% & 62.40\% & 65.48\% & -3.08\% & 90.76\% & 97.83\% & -7.07\% \\
    MGAN  & 49.26\% & 50.19\% & -0.93\% & 52.30\% & 54.81\% & -2.51\% & 84.03\% & 82.61\% & 1.42\% \\
    PRNet & 54.15\% & 54.72\% & -0.56\% & 59.92\% & 59.21\% & 0.71\% & 86.55\% & 89.13\% & -2.58\% \\
    \midrule
    \textbf{Average} & \textbf{26.78\%} & \textbf{27.36\%} & \textbf{-0.58\%} & \textbf{30.42\%} & \textbf{29.63\%} & \textbf{0.80\%} & \textbf{44.75\%} & \textbf{44.02\%} & \textbf{0.73\%} \\
    \bottomrule
\end{tabular}}
  \label{table8:contrast}
\end{table*}

Following the roadmap outlined in RQ2.1, we begin by comparing the overall miss rates of pedestrian detectors across different contrast levels, as presented in Table \ref{table8:contrast}. As explained in Section \ref{scenario}, we categorize driving scenarios into three contrast levels, with a higher level indicating greater contrast. However, we do not observe a consistent pattern in the results. Specifically, concerning light-skin and dark-skin pedestrians, we note a decrease in miss rates overall with increasing contrast. Nonetheless, this pattern is not observed across other demographic groups.

We then examine the shifts in fairness, as evidenced by the trends in EOD with increasing contrast, outlined in Table \ref{table8:contrast}. Our analysis reveals a consistent pattern: on average, detectors exhibit the most biased results (i.e., the largest absolute value of EOD) under level 2, while demonstrating the fairest outcomes under level 3.

Specifically, for gender, detectors achieve -1.63\% EOD under level 2 (most biased), -1.26\% under level 1, and 0.93\% under level 3 (fairest). For age, detectors achieve -21.05\% EOD under level 2 (most biased), -18.32\% under level 1, and 18.08\% under level 3 (fairest). For skin tone, detectors achieve 0.80\% EOD under level 2 (most biased), 0.73\% under level 1, and -0.58\% under level 3 (fairest).

This finding is further supported by the observation that under level 2, the highest number of detectors exhibit significant biases for gender, age, and skin tone, whereas under level 3, the fewest detectors show significant biases. Specifically, under level 2, five detectors demonstrate significant biases regarding gender, whereas under level 3, only one detector does so. Under level 2, eight detectors all demonstrate significant biases regarding age, whereas under level 3, seven detectors do so. Under level 2, four detectors show significant biases regarding skin tone, whereas under level 3, no detector exhibits significant biases.

\finding{\textbf{RQ2.2:} We classify driving scenarios into three contrast levels and observe that while there is not a clear pattern in the overall detection performance change with contrast variation, the highest contrast level consistently produces the fairest detection results across gender, age, and skin tone.}

\subsubsection{RQ2.3: Different weather conditions.} As decribed in Section \ref{scenario}, we consider two weather conditions: non-rainy and rainy. 

Table \ref{table8:non-rainy weather and rainy weather} presents the miss rates of eight pedestrian detectors under the two conditions, with statistically significant unfairness results shaded. Overall, the miss rate for each demographic group increases in rainy weather conditions. This escalation could be attributed to droplets covering the camera and disrupting the detectors. A follow-up question is 
whether rainy weather fairly increases the miss rate for different demographic groups.

\begin{table*}[!tp]
  \setlength{\tabcolsep}{2pt}
  \footnotesize
  \centering
  \caption{(RQ2.3) Miss rates and EOD of each pedestrian detector under non-rainy and rainy weather conditions. Statistically significant unfairness results are shaded. We find that rainy weather does not largely impact fairness regarding gender, age, and skin tone.}

  \scalebox{0.88}{ 
    \begin{tabular}{c|ccc|ccc}
    \toprule
    \multicolumn{7}{c}{\textbf{Gender}} \\
    \midrule
    \multirow{2}[4]{*}{\textbf{Detectors}} & \multicolumn{3}{c|}{\textbf{Non-rainy weather}} & \multicolumn{3}{c}{\textbf{Rainy weather}} \\
\cmidrule{2-7}          & \textbf{MR Male} & \textbf{MR Female} & \textbf{EOD} & \textbf{MR Male} & \textbf{MR Female} & \textbf{EOD} \\
    \midrule
    YOLOX & 9.58\% & 8.95\% & 0.63\% & 8.33\% & 5.17\% & 3.16\% \\
    RetinaNet & 10.62\% & 11.35\% & -0.74\% & 10.09\% & 8.05\% & 2.04\% \\
    Faster RCNN & 3.68\% & 3.29\% & 0.39\% & 4.39\% & 2.87\% & 1.51\% \\
    Cascade RCNN & 3.68\% & 3.48\% & 0.20\% & 4.39\% & 2.87\% & 1.51\% \\
    ALFNet & 23.52\% & 22.92\% & 0.60\% & 42.54\% & 40.80\% & 1.74\% \\
    CSP   & 25.29\% & 24.35\% & 0.94\% & 50.44\% & 44.25\% & 6.19\% \\
    MGAN  & 24.63\% & 25.26\% & -0.62\% & 50.88\% & 50.57\% & 0.30\% \\
    PRNet & 32.73\% & 31.64\% & 1.09\% & 51.75\% & 50.57\% & 1.18\% \\
    \midrule
    \textbf{Average} & \textbf{16.72\%} & \textbf{16.41\%} & \textbf{0.31\%} & \textbf{27.85\%} & \textbf{25.65\%} & \textbf{2.20\%} \\
    \midrule
    \multicolumn{7}{c}{\textbf{Age}} \\
    \midrule
    \multirow{2}[4]{*}{\textbf{Detectors}} & \multicolumn{3}{c|}{\textbf{Non-rainy weather}} & \multicolumn{3}{c}{\textbf{Rainy weather}} \\
\cmidrule{2-7}          & \textbf{MR Adult} & \textbf{MR Child} & \textbf{EOD} & \textbf{MR Adult} & \textbf{MR Child} & \textbf{EOD} \\
    \midrule
    YOLOX & \colorcell10.45\% & \colorcell39.95\% & \colorcell-29.50\% & \colorcell9.40\% & \colorcell30.00\% & \colorcell-20.60\% \\
    RetinaNet & \colorcell12.49\% & \colorcell40.65\% & \colorcell-28.16\% & \colorcell13.61\% & \colorcell40.00\% & \colorcell-26.39\% \\
    Faster RCNN & \colorcell4.36\% & \colorcell23.09\% & \colorcell-18.73\% & \colorcell4.86\% & \colorcell30.00\% & \colorcell-25.14\% \\
    Cascade RCNN & \colorcell4.53\% & \colorcell23.79\% & \colorcell-19.25\% & \colorcell4.38\% & \colorcell25.00\% & \colorcell-20.62\% \\
    ALFNet & \colorcell25.00\% & \colorcell47.34\% & \colorcell-22.34\% & \colorcell45.06\% & \colorcell75.00\% & \colorcell-29.94\% \\
    CSP   & \colorcell26.86\% & \colorcell44.11\% & \colorcell-17.25\% & 48.78\% & 65.00\% & -16.22\% \\
    MGAN  & \colorcell26.49\% & \colorcell41.11\% & \colorcell-14.62\% & 53.00\% & 65.00\% & -12.00\% \\
    PRNet & \colorcell33.19\% & \colorcell53.58\% & \colorcell-20.39\% & 53.32\% & 75.00\% & -21.68\% \\
    \midrule
    \textbf{Average} & \colorcell\textbf{17.92\%} & \colorcell\textbf{39.20\%} & \colorcell\textbf{-21.28\%} & \colorcell\textbf{29.05\%} & \colorcell\textbf{50.63\%} & \colorcell\textbf{-21.57\%} \\
    \midrule
    \multicolumn{7}{c}{\textbf{Skin Tone (LS: Light Skin, DS: Dark Skin)}} \\
    \midrule
    \multirow{2}[4]{*}{\textbf{Detectors}} & \multicolumn{3}{c|}{\textbf{Non-rainy weather}} & \multicolumn{3}{c}{\textbf{Rainy weather}} \\
\cmidrule{2-7}          & \textbf{MR LS} & \textbf{MR DS} & \textbf{EOD} & \textbf{MR LS} & \textbf{MR DS} & \textbf{EOD} \\
    \midrule
    YOLOX & 4.19\% & 3.47\% & 0.72\% & 16.94\% & 13.51\% & 3.42\% \\
    RetinaNet & \colorcell6.56\% & \colorcell3.81\% & \colorcell2.74\% & 21.77\% & 13.51\% & 8.26\% \\
    Faster RCNN & \colorcell4.59\% & \colorcell2.60\% & \colorcell1.99\% & 15.32\% & 10.81\% & 4.51\% \\
    Cascade RCNN & 3.93\% & 2.60\% & 1.33\% & 16.13\% & 8.11\% & 8.02\% \\
    ALFNet & 34.90\% & 36.92\% & -2.02\% & 66.13\% & 64.86\% & 1.26\% \\
    CSP   & 53.05\% & 57.02\% & -3.97\% & 79.03\% & 89.19\% & -10.16\% \\
    MGAN  & 44.68\% & 47.83\% & -3.15\% & 72.58\% & 78.38\% & -5.80\% \\
    PRNet & 51.99\% & 52.86\% & -0.87\% & 77.42\% & 81.08\% & -3.66\% \\
    \midrule
    \textbf{Average} & \textbf{25.49\%} & \textbf{25.89\%} & \textbf{-0.40\%} & \textbf{45.67\%} & \textbf{44.93\%} & \textbf{0.73\%} \\
    \bottomrule
    \end{tabular}}
  \label{table8:non-rainy weather and rainy weather}
\end{table*}

From Table \ref{table8:non-rainy weather and rainy weather}, we observe that rainy weather may potentially mitigate bias in pedestrian detectors. Specifically, under non-rainy conditions, all eight detectors exhibit significant bias against children, while under rainy conditions, three out of four pedestrian-specific detectors no longer show significant age bias any more. Similarly, under non-rainy conditions, two of eight detectors display significant bias against skin tone, whereas under rainy conditions, none do. However, the improvement in fairness due to rainy weather is marginal. The EOD difference between rainy and non-rainy conditions is 1.89\% for gender, -0.29\% for age, and 1.13\% for skin tone.

\finding{\textbf{RQ2.3: } Rainy weather conditions decrease overall detection performance to a large extent but have a subtle impact on the fairness of pedestrian detectors. Specifically, three out of four pedestrian-specific detectors no longer exhibit significant age bias under rainy weather. However, the improvement in fairness is marginal.}

\vspace{-5mm}
\section{Discussion}

\subsection{Fairness-performance Trade-off}\label{tradeoff}
It is widely acknowledged that fairness improvement usually comes at the cost of machine learning performance (e.g., accuracy) \cite{DBLP:conf/wsdm/GeZYPHHZ22,NEURIPS2019_373e4c5d,10.1145/3583561,10.1145/3468264.3468565}. Therefore, developers need to grapple with the challenge of optimizing ML performance without compromising fairness, encapsulating this tension as the ``fairness-performance trade-off.''

Nonetheless, we have observed that pedestrian detectors exhibiting greater fairness (i.e., lower absolute EOD values) can achieve superior performance (i.e., lower overall miss rates) under certain environmental conditions. For instance, in the results of RQ2.1, we find that during day time compared to night time, the eight pedestrian detectors achieve better detection results while simultaneously displaying reduced absolute EOD values related to age, gender, and skin tone. Similarly, the results of RQ2.2 reveal that the highest level of contrast consistently leads to improved fairness regarding gender, age, and skin tone, while also demonstrating the best overall detection performance in specific cases. These findings offer a positive counterpoint to previous fairness-performance trade-off theory, suggesting that we can enhance both detection performance and fairness by adjusting the brightness and contrast of captured images.

\subsection{Impact of Data Imbalance}
Imbalanced data distribution, particularly the underrepresentation of certain demographic groups such as children, may pose a threat to the validity of our findings. To evaluate the robustness of our results under varying data distributions, we perform a sensitivity analysis on age, where the representation of children is most limited and exhibits notable bias across all pedestrian detectors.

In this analysis, we systematically vary the proportion of adult samples by randomly sampling them at 80\%, 60\%, 40\%, and 20\% of their original size, while maintaining the number of samples for children constant. This creates four new data distributions with varying adult-to-child ratios. We then re-evaluate the age bias for each distribution. The results, presented in Table \ref{table11:sensitive analysis}, show that the observed age bias remains significant across all these distributions, indicating that our findings are robust even under substantial variations in the dataset's demographic composition.

\begin{table*}[!tp]
    \setlength{\tabcolsep}{2pt}
    \tiny
    \centering
    \caption{Sensitivity analysis of age bias under varying proportions of adult data. Statistically significant unfairness results are shaded. The analysis shows that the observed age bias remains significant across all distributions, indicating that our findings are robust even under substantial variations in the dataset's demographic composition.}
    \label{table11:sensitive analysis}
    \begin{tabular}{c|ccc|ccc|ccc|ccc}
    \toprule
    \multirow{2}{*}{Detectors} & \multicolumn{3}{c|}{Adult data proportion 80\%} & \multicolumn{3}{c|}{Adult data proportion 60\%} & \multicolumn{3}{c|}{Adult data proportion 40\%} & \multicolumn{3}{c}{Adult data proportion 20\%} \\ 
    \cmidrule(lr){2-4} \cmidrule(lr){5-7} \cmidrule(lr){8-10} \cmidrule(lr){11-13} 
                                & MR Adult & MR Child & EOD               & MR Adult & MR Child & EOD               & MR Adult & MR Child & EOD               & MR Adult & MR Child & EOD               \\ 
    \midrule
    YOLOX                        &\colorcell 12.68\%  &\colorcell 42.47\% &\colorcell -29.79\%                 &\colorcell 12.92\%  &\colorcell 42.47\% &\colorcell -29.55\%                 &\colorcell 12.73\%  &\colorcell 42.47\% &\colorcell -29.74\%                 &\colorcell 14.06\%  &\colorcell 42.47\% &\colorcell -28.41\%                 \\
    RetinaNet           &\colorcell 14.25\%  &\colorcell 44.33\% &\colorcell -30.08\%                 &\colorcell 14.31\%  &\colorcell 44.33\% &\colorcell -30.02\%                 &\colorcell 14.72\%  &\colorcell 44.33\% &\colorcell -29.61\%                 &\colorcell 15.07\%  &\colorcell 44.33\% &\colorcell -29.26\%                 \\
    Faster RCNN             &\colorcell 5.20 \%  &\colorcell 26.06\% &\colorcell -20.86\%                 &\colorcell 5.38 \%  &\colorcell 26.06\% &\colorcell -20.68\%                 &\colorcell 5.30 \%  &\colorcell 26.06\% &\colorcell -20.75\%                 &\colorcell 5.29\%   &\colorcell 26.06\% &\colorcell -20.77\%                 \\
    Cascade RCNN            &\colorcell 5.07 \%  &\colorcell 26.57\% &\colorcell -21.49\%                 &\colorcell 5.24 \%  &\colorcell 26.57\% &\colorcell -21.32\%                 &\colorcell 5.24 \%  &\colorcell 26.57\% &\colorcell -21.33\%                 &\colorcell 5.02\%   &\colorcell 26.57\% &\colorcell -21.55\%                 \\
    ALFNet               &\colorcell 36.35\%  &\colorcell 53.47\% &\colorcell -17.11\%                 &\colorcell 35.88\%  &\colorcell 53.47\% &\colorcell -17.58\%                 &\colorcell 36.27\%  &\colorcell 53.47\% &\colorcell -17.20\%                 &\colorcell 34.12\%  &\colorcell 53.47\% &\colorcell -19.35\%                 \\
    CSP             &\colorcell 37.51\%  &\colorcell 50.42\% &\colorcell -12.92\%                 &\colorcell 37.03\%  &\colorcell 50.42\% &\colorcell -13.40\%                 &\colorcell 37.35\%  &\colorcell 50.42\% &\colorcell -13.07\%                 &\colorcell 34.95\%  &\colorcell 50.42\% &\colorcell -15.47\%                 \\
    MGAN               &\colorcell 33.18\%  &\colorcell 46.53\% &\colorcell -13.35\%                 &\colorcell 32.94\%  &\colorcell 46.53\% &\colorcell -13.60\%                 &\colorcell 32.87\%  &\colorcell 46.53\% &\colorcell -13.66\%                 &\colorcell 30.65\%  &\colorcell 46.53\% &\colorcell -15.89\%                 \\
    PRNet           &\colorcell 44.40\%  &\colorcell 61.25\% &\colorcell -16.85\%                 &\colorcell 44.18\%  &\colorcell 61.25\% &\colorcell -17.08\%                 &\colorcell 44.05\%  &\colorcell 61.25\% &\colorcell -17.20\%                 &\colorcell 42.09\%  &\colorcell 61.25\% &\colorcell -19.17\                 \\ 
    \bottomrule
   \end{tabular}
\end{table*}

\subsection{Implications}
\subsubsection{Implication for researchers} Our empirical results offer valuable research opportunities for researchers. \textbf{1) Fairness improvement of autonomous driving systems via image editing.} Our findings reveal that brightness and contrast of images can impact the fairness of pedestrian detection. Thus, to improve fairness, a practical solution is to design specific post-processing image editing techniques to adjust contrast and brightness of captured input images, such as increasing brightness and contrast levels to dynamically counterbalance existing bias towards children and female pedestrians in low-brightness and low-contrast conditions. Moreover, previous discussions in Section \ref{tradeoff} show that these solutions also have the potential to improve detection performance.
\textbf{2) Real-time adaptive fairness improvement of autonomous driving systems.} 
Adapting software systems to increasingly dynamic environments poses significant challenges in software development, an area that has been an important task in SE research \cite{icseQureshiP09}. Our empirical findings contribute to this ongoing task by introducing a new challenge: designing real-time adaptive fairness improvement methods for autonomous driving systems. Specifically, our research reveals that autonomous driving systems exhibit greater biases under specific driving scenarios influenced by changes in brightness and contrast. Given the common occurrence of such changes in real-world driving conditions, there is an urgent need to develop specific methods to ensure the consistent fairness of autonomous driving systems in response to these environmental dynamics.
\textbf{3) Fairness improvement of autonomous driving systems via multi-objective optimization.} Given our finding that the fairness-performance trade-off may not hold in the context of pedestrian detection, researchers have the opportunity to develop multi-objective optimization strategies for model training, enabling the concurrent optimization of fairness and detection performance. At the same time, based on previous findings regarding brightness and contrast, a possible approach could involve integrating fairness constraints during model training by incorporating contrast-aware or brightness-aware regularization techniques. The model could penalize overly confident predictions in low-contrast or low-brightness conditions, encouraging it to treat predictions more cautiously in these scenarios. Additionally, augmenting data by enhancing these attributes in images of underrepresented groups (e.g., children in low-brightness conditions) could help the model generalize better. Synthetic image generation that varies brightness and contrast during training may further expose the model to diverse brightness and contrast conditions, potentially reducing bias. \textbf{4) Addressing inherent limitations for detecting small objects of pedestrian detectors. } For age bias, it is important to address the inherent limitations of pedestrian detectors, particularly their difficulty in accurately detecting small pedestrians. These implications underline the importance of factoring these elements into effective fairness improvement strategies. 
\textbf{5) Open datasets and annotations for fairness assessment.} There is a shortage of available datasets annotated with a wide range of demographic labels for pedestrian detection. To address this constraint, we have undertaken the task of manually annotating four datasets with information related to gender, age, and skin tone, and publicly released these annotated datasets. It is worth noting that there are additional demographic attributes that also need to be considered in the context of fairness, such as disability status and religion. We encourage researchers and practitioners to join us in the effort to promote fairness in pedestrian detection by creating and sharing datasets that encompass various demographic attributes.

\subsubsection{Implication for developers}
Fairness is a critical non-functional requirement for software applications, but our study demonstrates the existence of significant bias in state-of-the-art pedestrian detectors. It is imperative for developers to prioritize their efforts in this domain. Addressing these concerns goes beyond enhancing the quality of autonomous driving systems; it also serves to safeguard autonomous driving companies from ethical, reputational, financial, and legal repercussions that may arise in the event of violations of anti-discrimination laws. Moreover, it is crucial for developers of autonomous driving systems to consider the influence of various environmental factors on fairness during the development process.

\subsubsection{Implication for policy makers}  Governments have a role in raising awareness about the potential biases in autonomous driving systems. 
\textbf{1) Regulatory measures.} Autonomous driving systems play a crucial role in ensuring human safety. Our results reveal that fairness issues exist in modern pedestrian detection models. Therefore, it is essential for policy makers to enact regulations and standards that safeguard the rights of all individuals and address these concerns appropriately.
\textbf{2)~Enhanced protection for vulnerable pedestrians.} Policies should also aim to address safety issues for vulnerable pedestrians such as children and females, who are, as revealed by this study, more likely to be overlooked by the current detection models.

\subsection{Threats to Validity}
\noindent\textbf{Manual labeling.} The labeling process involves possible subjectivity, posing threats to the validity of the analysis. To mitigate this threat, two annotators and an arbitrator are involved in the labeling process: first each image is independently labeled by two annotators, and in cases where conflicts in labeling arise, we seek the expertise of an arbitrator to resolve such discrepancies and arrive at a consensus. The inter-rater agreement between the two annotators is high, which demonstrates the reliability of the labeling schema and procedure adopted herein.

\noindent\textbf{Selection of pedestrian detectors.} Our study is based on eight pedestrian detectors, which may lead to possible bias in this study. To mitigate this threat, we select representative pedestrian detectors based on two considerations. On one hand, the selected detectors are widely-studied in pedestrian detection and autonomous driving literature \cite{Cao2020FromHT,7780510,8099957,Yu_2020_CVPR}. On the other hand, the selected models cover the two typical types of pedestrian detection methods (i.e., one-stage detection and two-stage detection), ensuring a comprehensive representation of the techniques used in the field. 

\noindent\textbf{Selection of datasets.} The choice of datasets presents a potential threat to validity. Current literature on autonomous driving sometimes relies on automated techniques to simulate adverse weather conditions such as fog, rain, and snow. However, these synthetic images may not accurately reflect real-life conditions \cite{tseZhangHML22} (e.g., simply simulating fog by applying a set of filters over the original image, without considering the complex transformations that occur in actual road scenes under such conditions). Although real-world datasets for adverse weather like fog now exist~\cite{cvprBijelicGMKRDH20}, accurately annotating demographic attributes of pedestrians under foggy conditions remains challenging in the context of this study. These weather effects can obscure important features such as skin tone and blur physical outlines, making it difficult to discern attributes like gender and age. To address these concerns, we utilize four real-world datasets extensively explored in autonomous driving research, comprising a total of 8,311 images. Additionally, we have augmented these datasets with 16,070 gender labels, 20,115 age labels, and 3,513 skin tone labels. The extensive scale of data ensures the reliability of our results and mitigates the limitations associated with synthetic images.

\noindent\textbf{Selection of evaluation measures.} The fairness measures that we employ may introduce potential limitations. To mitigate this concern, we consider three commonly-used fairness measures: SPD, EOD, and AOD. Upon careful discussion, we conclude that AOD may not be suitable for our study (Section \ref{evalution}), leading us to focus on SPD and EOD. We demonstrate that SPD and EOD lead to equivalent observations for pedestrian detection. Ultimately, we opt to use EOD (i.e., SPD) for our analysis, as it involves the comparison of the miss rates of different demographic groups. This ensures a consistent evaluation with the current pedestrian detection research, as miss rate is the most widely-adopted metric for measuring the performance of pedestrian detectors in the literature~\cite{Brandao2019AgeAG,Kogure2022AgeSN}.

\noindent\textbf{Selection of sensitive attributes.} The selection of sensitive attributes may introduce potential limitations to our study. To mitigate this concern, we focus on attributes that are most identifiable in autonomous driving images and are recognized as the three most extensively studied sensitive attributes in the fairness literature: gender, age, and skin tone \cite{DBcorrabs220707068}. While other sensitive attributes, such as disability status, precise age determinations for young and elderly individuals, or nationality, are relevant to fairness considerations, they present significant challenges for consistent identification from typical vehicle-mounted camera data due to ambiguous feature judgment. In contrast, gender, age (adult and child), and skin tone are more readily observable characteristics in such images, allowing for more reliable annotations.

\section{Conclusion}\label{conclusion}
This paper presents the first comprehensive study on fairness testing of eight state-of-the-art pedestrian detectors, using four widely-studied testing datasets. We investigate the fairness aspects of these detectors regarding gender, age, and skin tone. Furthermore, we conduct an in-depth analysis of fairness in various driving scenarios. Our findings reveal significant bias in the current pedestrian detectors, particularly towards children. Additionally, pedestrian detectors demonstrate significant gender biases during night time, potentially exacerbating the prevalent societal issue of female safety concerns during nighttime out. Regarding skin tone, we observe balanced detection performance for light-skin and dark-skin individuals overall. As part of our contribution, we publicly release large-scale real-world pedestrian detection datasets with gender, age, and skin tone labels. These datasets aim to facilitate the future fairness research in autonomous driving.
The insights gained in this study can pave the way for more fair and unbiased autonomous driving systems in the future.

\section{Data Availability}
Our GitHub repository \cite{githublink} contains datasets, sensitive attribute labels, scripts, and results of this work to facilitate replication and extension.

\section*{Acknowledgement}
This work was supported by the National Natural Science Foundation of China under the grant number 62325201, as well as by the Center for Data Space Technology and Systems at Peking University. Xinyue Li, Ying Zhang, and Xuanzhe Liu are also affiliated with the Key Lab of High Confidence Software Technologies (Peking University), Ministry of Education, Beijing, China.

\bibliographystyle{ACM-Reference-Format}
\bibliography{fairness-bib}
\end{document}